\documentclass[12pt,a4paper]{article}

\usepackage{amsfonts}

\newcommand{\Io}{{\mathbb I}}
\newcommand{\Ro}{{\mathbb R}}

\newcommand{\Eo}{{\mathbb E}}
\newcommand{\Fo}{{\mathbb F}}

\newcommand{\be}{\begin{eqnarray}}
\newcommand{\ee}{\end{eqnarray}}

\newcommand{\email}[1]{{\small E-mail: #1}}

\newtheorem{proposition}{Proposition}
\newtheorem{theorem}{Theorem}

\newtheorem{lemma}{Lemma}
\newtheorem{definition}{Definition}
\newenvironment{proof}
{\par\noindent {\bf Proof}}
{\par\strut\hfill$\square$\par\vskip 0.5cm}

\begin{document}

\title{Estimators, escort probabilities, and\\
 $\phi$-exponential  families in\\
statistical physics}
\author{
Jan Naudts\\
\small Departement Natuurkunde, Universiteit Antwerpen UIA,\\
\small Universiteitsplein 1, 2610 Antwerpen, Belgium\\
\email {Jan.Naudts@ua.ac.be}
}

\date{}

\maketitle

\begin{abstract}
The lower bound of Cramer and Rao is generalized to pairs of families
of probability distributions, one of which is escort to the other.
This bound is optimal for certain families, called $\phi$-exponential in the paper.
Their dual structure is explored.
\end{abstract}

\section{Introduction}

Aim of this paper is to translate some new results of statistical physics
into the language of statistics. It is well-known that the
exponential family of probability distribution functions (pdfs)
plays a central role in statistical physics. When Gibbs introduced
the canonical ensemble in 1901 \cite {GJW01} he postulated a
distribution of energies $E$ of the form
\be
p(E)=\exp(G-\beta E)
\label {bg}
\ee
where $G$ is a normalization constant and where the control parameter $\beta$
is the inverse temperature.
Only recently \cite {TC88}, a proposal was made to replace (\ref {bg})
by a more general family of pdfs. The resulting domain of research is known under the name
of Tsallis' thermostatistics. Some of the
pdfs of Tsallis' thermostatistics are known in statistics under the name of
Amari's $\alpha$-family \cite {AS85}. The latter have been introduced
in the context of geometry of statistical manifolds \cite {MR93}.
The appearance of the same family of pdfs in both domains is not accidental.
The apparent link between both domains is clarified in the present paper.

The new notion introduced in Tsallis' thermostatistics is that of
pairs of families of pdfs, one of which is {\sl escort} of the other
\cite {BS93}. Some basic concepts of statistics can be generalized
by replacing at well-chosen places the pdf by its escort.
In particular, we show in the next section how to generalize
Fisher's information and, correspondingly, how to generalize the
well-known lower bound of Cramer and Rao.
Section 3 studies the statistical manifold of a family for which
there exists an escort family satisfying the condition under which
the generalized Cramer-Rao bound is optimal.
This optimizing family has an affine geometry. Since this is usually
the characteristic property of an exponential family
a generalization of the latter seems indicated.

Section 4 shows how a strictly positive non-decreasing function $\phi$ of $\Ro_+$
determines a function which shares some properties with the natural logarithm
and therefore is called below a $\phi$-logarithm. The inverse
function is called the $\phi$-exponential. In Section 5
it is used to define the $\phi$-exponential family in the
obvious way, by replacing the exponential function $\exp$ by the
$\phi$-exponential function. The standard exponential family is
then recovered by the choice $\phi(x)=x$, the $\alpha$-family of Amari
by $\phi(x)=x^{(1+\alpha)/2}$, the equilibrium pdfs of Tsallis' thermostatistics
by the choice $\phi(x)=x^q$.

The next three sections are used to establish the dual parametrization of
the $\phi$-exponential family and to discover the role of entropy functionals.
Section 6 introduces a divergence of the Bregman type. In Section 7 it is used
to prove the existence of an information function (or entropy functional)
which is maximized by the $\phi$-exponential pdfs. Section 8 introduces
dual parameters --- in statistical physics these are energy and temperature.
The paper ends with a short discussion in Section 9.

There have been already some attempts to study Tsallis' thermostatistics
from a geometrical point of view. Trasarti-Battistoni \cite {TBR02} conjectured a
deep connection between non-extensivity and geometry. He also gives general references
to the use of geometric ideas in statistical physics. 
Several authors \cite {AS98,TC98,SM98} have introduced a divergence
belonging to Csisz\'ar's class of f-divergences, which leads to a generalization
of the Fisher information metric adapted to the context of Tsallis' thermostatistics.
The relation with the present work is unclear since here the geometry is determined
by a divergence of the Bregman type. Also the recent work of
Abe \cite {AS03} seems to be unrelated.

\section{Estimators and escort pdfs}

Fix a measure space $\Omega,\mu$.
Let ${\cal M}_1(\mu)$ denote the convex set of all probability distribution
functions (pdfs) $p$ normalized w.r.t.~$\mu$
\be
\int_{\Omega}{\rm d}\mu(x)\,p(x)=1.
\label {norm}
\ee
Expectations w.r.t.~$p$ are denoted $\Eo_p$
\be
\Eo_p f=\int_{\Omega}{\rm d}\mu(x)\,p(x)f(x).
\ee

Fix an open domain $D$ of $\Ro^n$. Consider a family of
pdfs $p_\theta$, parametrized with $\theta$ in $D$. The notation
$\Eo_\theta$ will be used instead of $\Eo_{p_\theta}$.
Simultaneously, a second family of pdfs $(P_\theta)_{\theta\in D}$
is considered. It is called the {\sl escort family}.
The notation $\Fo_\theta$ will be used instead of $\Eo_{P_\theta}$.

Recall that the Fisher information is given by
\be
I_{kl}(\theta)
&=&\Eo_\theta
\left(\frac{\partial\,}{\partial \theta^k}\log(p_\theta)\right)
\left(\frac{\partial\,}{\partial \theta^l}\log(p_\theta)\right)\cr
&=&
\int_{\Omega}{\rm d}\mu(x)\,\frac{1}{p_\theta(x)}
\frac{\partial p_\theta}{\partial \theta^k}
\frac{\partial p_\theta}{\partial \theta^l}.
\label {fisher}
\ee
A generalization, involving the two families of pdfs, is
\be
g_{kl}(\theta)
&=&\int_{\Omega}{\rm d}\mu(x)\,\frac{1}{P_\theta(x)}
\frac{\partial p_\theta}{\partial \theta^k}
\frac{\partial p_\theta}{\partial \theta^l}.
\label {gdef}
\ee
Clearly, the expression coincides with (\ref {fisher}) if $P_\theta=p_\theta$.

The following definition is a slight generalization of the usual definition
of an unbiased estimator.

\begin{definition}
An {\sl estimator} of the family $(p_\theta)_{\theta\in D}$
is a vector of random variables $c_k$ with the property that there
exists a function $F$ such that
\be
\Eo_\theta c_k=\frac{\partial\,}{\partial\theta^k}F(\theta),
\qquad k=1,\cdots,n.
\ee
The function $F$ will be called the {\sl scale function} of the estimator.
\end{definition}
The estimator is unbiased if $F(\theta)=\frac 12\theta_k\theta^k$
so that $\Eo_\theta c_k=\theta_k$.
The well-known lower bound of Cramer and Rao can be written as
\be
\frac{u^ku^l\left[\Eo_\theta c_kc_l-\big(\Eo_\theta c_k\big)\big(\Eo_\theta c_l\big)\right]}
{\left[u^kv^l\frac{\partial^2F}{\partial\theta^k\partial\theta^l}\right]^2}
\ge
\frac{1}{v^kv^lI_{kl}(\theta)},
\label {CramerRao}
\ee
for arbitrary $u$ and $v$ in $\Ro^n$.

A similar lower bound, involving the information matrix $g_{kl}$ instead of Fisher's $I_{kl}$,
is now formulated.

\begin{theorem}
Let be given two families of pdfs $(p_\theta)_{\theta\in D}$ and
$(P_\theta)_{\theta\in D}$ and corresponding expectations $\Eo_\theta$ and $\Fo_\theta$.
Let $c$ be an estimator of $(p_\theta)_{\theta\in D}$,
with scale function $F$.
Assume the regularity condition
\be
\Fo_\theta \frac{1}{P_\theta(x)}\frac{\partial\,}{\partial\theta^k}p_\theta(x)=0
\label {reg}
\ee
holds.
Let $g_{kl}(\theta)$ be the information matrix introduced before.
Then, for all $u$ and $v$ in $\Ro^n$ is
\be
\frac{u^ku^l\left[\Fo_\theta c_kc_l-\big(\Fo_\theta c_k\big)\big(\Fo_\theta c_l\big)\right]}
{\left[u^kv^l\frac{\partial^2\,}{\partial\theta^l\partial\theta^k}F(\theta)\right]^2}
\ge
\frac{1}{v^kv^lg_{kl}(\theta)}.
\label {lb}
\ee
The bound is optimal (in the sense that equality holds whenever $u=v$)
if there exist a normalization function $Z>0$ 
and a function $G$ such that
\be
\frac{\partial\,}{\partial\theta^k}p_\theta(x)
=Z(\theta)P_\theta(x)\frac{\partial\,}{\partial\theta^k}\big[G(\theta)-\theta^lc_l(x)\big]
\label {escort}
\ee
holds for all $k$ in $[1..m]$, for all $\theta\in D$, and for $\mu$-almost all $x$.
In that case, $c$ is an estimator of $(P_\theta)_{\theta\in D}$ with
scale function $G$
\be
\Fo_\theta c_k=\frac{\partial G}{\partial\theta^k}.
\ee
\end{theorem}

\begin{proof}
Let
\be
X_k=\frac{1}{P_\theta}\frac{\partial\,}{\partial\theta^k}p_\theta
\qquad\hbox{ and }\quad
Y_k=c_k-\Fo_\theta c_k.
\ee
From Schwartz's inequality follows
\be
\left(\Fo_\theta u^kY_kv^lX_l\right)^2
&\le&\left(\Fo_\theta u^k Y_ku^lY_l\right)
\left(\Fo_\theta v^kX_k v^lX_l\right).
\ee
The l.h.s.~equals, using (\ref {reg}),
\be
\left(\Fo_\theta u^kY_kv^lX_l\right)^2
&=&\left(u^kv^l\frac{\partial\,}{\partial\theta^l}\Eo_\theta c_k\right)^2\cr
&=&\left(u^kv^l\frac{\partial^2\,}{\partial\theta^l\partial\theta^k}F(\theta)\right)^2.
\ee
The first factor of the r.h.s.~equals
\be
\Fo_\theta u^k Y_ku^lY_l
&=&u^ku^l\left[\Fo_\theta c_kc_l-\big(\Fo_\theta c_k\big)\big(\Fo_\theta c_l\big)\right].
\ee
The second factor of the r.h.s.~equals
\be
\Fo_\theta v^kX_k v^lX_l
&=&v^kv^lg_{kl}(\theta).
\ee
This proves (\ref {lb}).

Assume now that (\ref {escort}) holds. Combining it with the regularity condition
(\ref {reg}) shows that $c$ is an estimator for the
escort family, with scaling function $G$. This makes it possible to write
(\ref {escort}) as
\be
\frac{1}{Z(\theta)P_\theta(x)}
\frac{\partial\,}{\partial\theta^k}p_\theta(x)
&=&\Fo_\theta c_k-c_k(x).
\label {score}
\ee
In this way one obtains
\be
u^ku^l\left[\Fo_\theta c_kc_l-\big(\Fo_\theta c_k\big)\big(\Fo_\theta c_l\big)\right]
&=&\frac{u^ku^lg_{kl}(\theta)}{Z(\theta)^2}.
\label {pr1}
\ee
On the other hand is
\be
\frac{\partial^2\,}{\partial\theta^l\partial\theta^k}F(\theta)
&=&\frac{\partial\,}{\partial\theta^l}\Eo_\theta c_k\cr
&=&\int_{\Omega}{\rm d}\mu(x)\,\frac{\partial p_\theta}{\partial\theta^l}(x)c_k(x)\cr
&=&Z(\theta)\int_{\Omega}{\rm d}\mu(x)\,P_\theta(x)c_k(x)
\frac{\partial\,}{\partial\theta^k}\big[G(\theta)-\theta^lc_l(x)\big]\cr
&=&-Z(\theta)\left[\Fo_\theta c_kc_l-\big(\Fo_\theta c_k\big)\big(\Fo_\theta c_l\big)\right].
\label {Fderiv}
\ee
Together with (\ref {pr1}) this shows equality in (\ref {lb}) whenever $u=v$.

\end{proof}

It is not investigated whether (\ref {escort}) is a necessary condition.
For practical application of the lower bound one has to assume that
$c$ is also an estimator of the escort family $(P_\theta)_{\theta\in D}$,
with scale function $G$. The previous proposition shows that this
is automatically the case when (\ref {escort}) is satisfied.

\paragraph {Example 1}
Let $\mu$ be the Lebesgue measure restricted to $[0,+\infty)$ and let
\be
p_\theta(x)=\frac 2\theta\left[1-\frac x{\theta}\right]_+
\ee
with $\theta>0$ and $[u]_+=\max\{u,0\}$.
The Fisher information $I(\theta)$ is divergent.
Hence, the usual lower bound of Cramer and Rao is useless.

Consider now the escort family
\be
P_\theta(x)=\frac 1\theta e^{-x/\theta}.
\ee
Then one calculates
\be
g(\theta)&=&\frac 4{\theta^2}(5e-13).
\ee
This fixes the r.h.s.~of the inequality (\ref {lb}).

Let us estimate $\theta$ via its first moment, with $c(x)=3x$.
One has $\Eo_\theta c=\theta$, $\Eo_\theta c^2=(3/2)\theta^2$, $F(\theta)=\theta^2/2$,
$\Fo c=3\theta$ and $\Fo c^2=18\theta^2$. Then (\ref {lb})
boils down to
\be
\Fo c^2-\big(\Fo c\big)^2
=9\theta^2\ge \frac 1{4(5e-13)}\theta^2\simeq 0.4 \,\theta^2.
\ee

\section{Statistical manifold}

The well-known example of a family with optimal estimator
is the exponential family
\be
p_\theta(x)=\exp\big(G(\theta)-\theta^kc_k(x)\big)
\ee
with
\be
G(\theta)=-\log\int_{\Omega}{\rm d}\mu(x)\,e^{-\theta^kc_k(x)}.
\ee
One sees immediately that
\be
\frac{\partial\,}{\partial\theta^k}p_\theta(x)
=p_\theta(x)\left(\frac{\partial\,}{\partial\theta^k}G(\theta)-c_k(x)\right),
\ee
which is (\ref {score}) with $Z(\theta)$ identically 1 and the escort pdf $P_\theta$
equal to $p_\theta$. This example motivates also the geometric
interpretation of (\ref {escort}), in the form (\ref {score}), as a linear map between tangent planes.
The score variables $\partial \log p_\theta/\partial \theta^k$ of the
standard statistical manifold are replaced by the variables
\be
\frac{1}{P_\theta(x)}
\frac{\partial\,}{\partial\theta^k}p_\theta(x).
\ee
They are tangent vectors of the concave function $G(\theta)-\theta^lc_l$.
The metric tensor of the latter function is a constant random variable. The
geometry of the manifold of random variables
$\big(G(\theta)-\theta^lc_l\big)_{\theta\in D}$ is transferred onto the family of pdfs
$\big(p_\theta\big)_{\theta\in D}$.

Note that the score variables have vanishing expectation $\Fo_\theta$.
It is now obvious to define an inner product of random variables by
\be
\langle A,B\rangle_\theta=\Fo_\theta AB.
\ee
Then one has
\be
\left\langle 
\frac{1}{P_\theta}\frac{\partial p_\theta}{\partial\theta^k},
\frac{1}{P_\theta}\frac{\partial p_\theta}{\partial\theta^l}
\right\rangle_\theta
&=&g_{kl}(\theta).
\ee
Let $g^{kl}(\theta)$ denote the inverse of $g_{kl}(\theta)$ (assume it exists).
Then a projection operator $\pi_\theta$ onto the orthogonal complement of
the tangent plane is defined by
\be
\pi_\theta A=A-g^{kl}\left\langle 
\frac{1}{P_\theta}\frac{\partial p_\theta}{\partial\theta^k},A\right\rangle_\theta
\frac{1}{P_\theta}\frac{\partial p_\theta}{\partial\theta^l}
-\Fo_\theta A.
\ee
If (\ref {escort}) is satisfied, then
\be
\pi_\theta \frac{\partial\,}{\partial\theta^l}
\frac{1}{P_\theta}\frac{\partial p_\theta}{\partial\theta^k}
&=&\pi_\theta\left[\frac{\partial Z}{\partial\theta^l}\left(\Fo_\theta c_k-c_k\right)
+Z(\theta)\frac{\partial^2 G}{\partial\theta^k\partial\theta^l}\right]\cr
&=&\frac{\partial Z}{\partial\theta^l}\left[
\Fo_\theta c_k-c_k+g^{lm}(\theta)\langle \frac{1}{P_\theta}
\frac{\partial p_\theta}{\partial\theta^l},c_k\rangle_\theta
\frac{1}{P_\theta}\frac{\partial p_\theta}{\partial\theta^m}
\right]\cr
&=&\frac{\partial Z}{\partial\theta^l}\left[
\Fo_\theta c_k-c_k-\frac{1}{Z(\theta)P_\theta}\frac{\partial p_\theta}{\partial\theta^k}\right]\cr
&=&0.
\ee
This follows also immediately from
\be
\frac{\partial\,}{\partial\theta^l}
\frac{1}{P_\theta}\frac{\partial p_\theta}{\partial\theta^k}
=\frac{1}{Z(\theta)}\frac{\partial Z}{\partial \theta^l}\frac{1}{P_\theta}\frac{\partial p_\theta}{\partial\theta^k}
+Z(\theta)\frac{\partial^2 G}{\partial\theta^k\theta^l}.
\ee
That the derivatives of the score variables are linear combinations
of the score variables and the constant random variable is usually
the characteristic feature of the exponential family.
This is a motivation to introduce a generalized notion of exponential family.

\section{$\phi$-logarithms and $\phi$-exponentials}

In the next section the notion of exponential family is generalized
to a rather large class of families of pdfs. This is done by replacing
the exponential function by some other function satisfying a minimal number
of requirements. The latter function will be called a deformed exponential
and will be denoted $\exp_\phi$. This has the advantage that the resulting
expressions look very familiar, resembling those of the exponential family.

Fix an increasing function $\phi$ of $[0,+\infty)$, strictly positive on $(0,+\infty)$.
It is used to define the $\phi$-logarithm $\ln_\phi$ by
\be
\ln_\phi(u)=\int_1^u{\rm d}v\,\frac{1}{\phi(v)}, \qquad u>0.
\ee
Clearly, $\ln_\phi$ is a concave function which is
negative on $(0,1)$ and positive on $(1,+\infty)$.
The inverse of the function $\ln_\phi$ is denoted $\exp_\phi$. It is
defined on the range of $\ln_\phi$. The definition can be extended to
all of $\Ro$ by putting $\exp_\phi(u)=0$ if $u$ is too small and $\exp_\phi=+\infty$
if $u$ is too large. In case $\phi(u)=u$ for all $u$ then $\ln_\phi$
coincides with the natural logarithm and $\exp_\phi$ coincides with
the exponential function.

Given $\phi$, introduce a function $\psi$ of $\Ro$ by
\be
\psi(u)
&=&\phi\big(\exp_\phi(u)\big)\quad\hbox{ if }u\hbox{ is in the range of }\ln_\phi\cr
&=&0\quad\hbox{ if }u\hbox{ is too small}\cr
&=&+\infty\quad\hbox{ if }u\hbox{ is too large}.
\label {psidef}
\ee
Clearly is $\phi(u)=\psi(\ln_\phi(u))$ for all $u>0$.

\begin{proposition}
One has for all $u$ in $\Ro$
\be
0\le \exp_\phi(u)&=&1+\int_0^u{\rm d}v\,\psi(v)\cr
&=&\int_{-\infty}^u{\rm d}v\,\psi(v)\le +\infty.
\label {lem1}
\ee
\end{proposition}

\begin{proof}
First consider the case that $[0,u)$ belongs to the range of $\ln_\phi$.
Then a substitution of integration variables
$v=\ln_\phi(w)$ is possible. One finds, using ${\rm d}v/{\rm d}w=1/\phi(w)$
and $\psi(v)=\phi\big(\exp_\phi(v)\big)=\phi(w)$,
\be
\int_0^u{\rm d}v\,\psi(v)
&=&\int_1^{\exp_\phi(u)}{\rm d}w\cr
&=&\exp_\phi(u)-1.
\ee
Using $\exp_\phi(-\infty)=0$ one concludes (\ref {lem1}).

In case $M=\sup_v\ln_\phi(v)$ is finite and $u\ge M$ then $\psi(v)=+\infty$ for $v\in[M,u]$.
One has
\be
\int_0^u{\rm d}v\,\psi(v)
&\ge&\int_0^M{\rm d}v\,\psi(v)\cr
&=&\int_1^{+\infty}{\rm d}w\cr
&=&+\infty.
\ee
But also the l.h.s.~of (\ref {lem1}) is infinite. Hence the equality holds.

Finally, if $m=\inf_v\ln_\phi(v)$ is finite and $u\le m$ then
$\psi(v)=0$ holds for $v\le m$. Hence
\be
\int_0^u{\rm d}v\,\psi(v)
&=&\int_0^m{\rm d}v\,\psi(v)\cr
&=&\int_1^0{\rm d}w\cr
&=&-1.
\ee
This ends the proof.
\end{proof}

\begin{proposition}
The function $\exp_\phi$ is continuous on the open interval of points
where it does not diverge.
\end{proposition}

\begin{proof}
Let $m$ and $M$ be as in the proof of the previous proposition.
Then $\exp_\phi$ is differentiable on $(m,M)$.
If $m=-\infty$ this ends the proof. If $m$ is finite
then it suffices to verify that $\exp_\phi(u)$ is
continuous in $u=m$. But this is straightforward.

\end{proof}

\paragraph {Example 2}
Let $\phi(u)=u^q$ with $q>0$. This function
is increasing and strictly positive on $(0,+\infty)$. Hence,
it defines a $\phi$-logarithm which will be denoted $\ln_q$
and is given by
\be
\ln_q(u)
&=&\int_1^u{\rm d}v\,\frac{1}{v^q}\cr
&=&\frac{u^{1-q}-1}{1-q}\qquad \hbox{ if }q\not=1\cr
&=&\log(u)\qquad \hbox{ if }q=1.
\ee
This deformed logarithm has been introduced
in the context of nonextensive statistical physics in \cite {TC94}.
The inverse function is denoted $\exp_q$ and is given by
\be
\exp_q(u)=\big[1+(1-q)u]_+^{1/(1-q)}.
\ee
The function $\psi$ is then given by
\be
\psi(u)=\big[1+(1-q)u]_+^{q/(1-q)}.
\ee

\paragraph {Example 3}
Let $\phi(x)=\lceil x\rceil$, the smallest integer not smaller than $x$.
This piecewise constant function
is increasing and strictly positive on $(0,+\infty)$. Hence,
$\ln_\phi$ is piecewise linear. 
The function $\psi$ is given by
\be
\psi(x)
&=&0\qquad\qquad\hbox{ if } x\le -1\cr
&=&\phi(1+x)\qquad\hbox{ otherwise}.
\label {egpsi}
\ee
The $\phi$-exponential $\exp_\phi$
is also piecewise linear and satisfies
\be
\exp_\phi(x)=0\qquad x\le -1.
\ee

\section{The $\phi$-exponential family}

Let $\phi$ be given as in the previous section.
Fix a measure space $\Omega,\mu$ and a set of random variables $c_k,k=1,\cdots,n$.
The $\phi$-exponential family of pdfs
$\big(p_\theta\big)_{\theta\in D}$ is defined by
\be
p_\theta(x)=\exp_\phi\big(G(\theta)-\theta^kc_k(x)\big).
\label {phiexp}
\ee
The domain $D$ is an open set of $\theta$ for which $G(\theta)$
exists such that (\ref {phiexp}) is properly normalized, i.e.~$p_\theta\in{\cal M}_1(\mu)$.
The distributions (\ref {phiexp}) are the equilibrium pdfs of generalized thermostatistics
as introduced in \cite {NJ04,NJ04b}.

\begin{proposition}
The function $G(\theta)$ is concave on $D$.
\end{proposition}

\begin{proof}
Assume $\theta$, $\eta$ and $\lambda\theta+(1-\lambda)\eta$ in $D$ for some $\lambda$ in $[0,1]$.
Then, using convexity of $\exp_\phi$,
\be
& &\exp_\phi\big(\lambda G(\theta)+(1-\lambda)G(\eta)-\big[\lambda\theta^k+(1-\lambda)\eta^k\big]c_k(x)\big)\cr
& &\le \lambda p_\theta(x)+(1-\lambda)p_\eta(x).
\ee
Hence
\be
\int_{\Ro^n}{\rm d}\mu(x)\,\exp_\phi\left(\lambda G(\theta)+(1-\lambda)G(\eta)
-\big[\lambda\theta^k+(1-\lambda)\eta k\big]c_k(x)\right)
\le 1.
\ee
Since $\exp_\phi$ is increasing one concludes that
\be
\lambda G(\theta)+(1-\lambda)G(\eta)\le G(\lambda \theta+(1-\lambda)\eta).
\ee
Hence $G$ is concave.
\end{proof}

\begin{proposition}
Let $\psi$ be determined by $\phi$ via (\ref {psidef}).
If the integral
\be
Z(\theta)=\int_{\Omega}{\rm d}\mu(x)\,\psi\big(G(\theta)-\theta^kc_k(x)\big)
\ee
converges for all $\theta\in D$, then $\big(p_\theta\big)_{\theta\in D}$
has an escort family $\big(P_\theta\big)_{\theta\in D}$, given by
\be
P_\theta(x)
&=&\frac{1}{Z(\theta)}\phi\big(p_\theta(x)\big)\qquad\hbox{ if } p_\theta(x)>0\cr
&=&0\qquad\hbox{ otherwise}.
\ee
Condition (\ref {escort}) is satisfied.
\end{proposition}

\begin{proof}
One has
\be
\phi\big(p_\theta(x)\big)
&=&\phi\big(\exp_\phi\big(G(\theta)-\theta^kc_k(x)\big)\big)\cr
&=&\psi\big(G(\theta)-\theta^kc_k(x)\big).
\ee
Because $\phi\big(p_\theta(x)\big)$ cannot be zero for $\mu$-almost all $x$
one concludes that $Z(\theta)>0$ and that $P_\theta$ is properly normalized.

From the properties of the function $\exp_\phi$ follows
immediately that
\be
\frac{\partial\,}{\partial\theta^l}p_\theta(x)
&=&\psi\big(G(\theta)-\theta^kc_k(x)\big)
\frac{\partial\,}{\partial\theta^l}\big(G(\theta)-\theta^mc_m(x)\big)\cr
&=&Z(\theta)P_\theta(x)
\frac{\partial\,}{\partial\theta^l}\big(G(\theta)-\theta^mc_m(x)\big).
\ee
This proves that $\big(P_\theta\big)_{\theta\in D}$ satisfies (\ref {escort}).

\end{proof}

\paragraph {Example 2 continued}
Let $\phi(u)=u^q$ as in Example 2 above.
The pdfs $p_\theta$ are given by
\be
p_\theta(x)&=&\left[1+(1-q)\big(G(\theta)-\theta^kc_k(x)\big)\right]_+^{1/(1-q)},
\ee
for $\theta$ in a suitable domain $D$.
The escort probabilities are
\be
P_\theta(x)&=&\frac{1}{Z(\theta)}\left[1+(1-q)\big(G(\theta)-\theta^kc_k(x)\big)\right]_+^{q/(1-q)}
\ee
with
\be
Z(\theta)&=&\int_{\Omega}{\rm d}\mu(x)\,
\left[1+(1-q)\big(G(\theta)-\theta^kc_k(x)\big)\right]_+^{q/(1-q)}
\ee
(assuming convergence of these integrals).
The family $\big(p_\theta\big)_{\theta\in D}$ coincides with
Amari's $\alpha$-family \cite {AS85}, with $\alpha$ given by $\alpha=2q-1$.

\paragraph {Example 1 continued}
Example 1 is the $q=0$-limit of example 2. Let $\phi(u)=1$ for all $u>0$.
Then
\be
\ln_\phi(u)&=&u-1\cr
\exp_\phi(u)&=&[1+u]_+\cr
\psi(u)&=&1\qquad\hbox{ if }u > -1;\cr
&=&0\qquad\hbox{ otherwise}.
\ee
One has
\be
p_\theta(x)&=&\frac 2\theta \left[1-\frac x\theta\right]_+\cr
&=&\exp_\phi\left(\frac 2\theta-1-\frac {2x}{\theta^2}\right).
\ee
This is a $\phi$-exponential family with parameter $\Theta=1/\theta^2$,
estimator $c(x)=2x$ and scale function $G(\Theta)=2\sqrt\Theta$.
The escort probabilities, making inequality (\ref {lb}) optimally
satisfied, are given by
\be
P_\Theta(x)&=&\frac 1\theta\Io_{0\le x\le \theta}.
\ee
The information matrix $g(\Theta)$ equals $\theta^4/3$.
Further is $\Fo_\Theta c=\theta$ and $\Fo_\Theta c^2=4\theta^2/3$
and
\be
\frac {\partial\,}{\partial \Theta}F(\Theta)=\Eo_\Theta c=2\theta/3=2/3\sqrt\Theta.
\ee
It is now straightforward to verify that the inequality (\ref {lb})
is optimally satisfied.

\section{Divergences}

Divergences of the Bregman type are needed for what follows.
In the form given below they have been introduced in \cite {NJ03}.

Fix a strictly positive increasing function $\phi$ of $[0,+\infty)$.
Introduce
\be
D_\phi(p||p')
&=&\int_{\Omega}{\rm d}\mu(x)\,\int_{p'(x)}^{p(x)}{\rm d}u\,
\left[\ln_\phi(u)-\ln_\phi(p'(x))\right].
\ee
$D_\phi(p||p')\ge 0$ follows because $\ln_\phi$ is an increasing function.
Also convexity in the first argument follows because $\ln_\phi$ is an increasing function.

Let $\big(p_\theta\big)_{\theta\in D}$ be $\phi$-exponential. Then infinitesimal variation
of the divergence $D_\phi(p||p')$ reproduces the metric tensor $g_{kl}(\theta)$, up to
a scalar function. Indeed, one has
\be
\frac{\partial\,}{\partial\theta^k}D_\phi(p_\theta||p_\eta)\big|_{\eta=\theta}&=&0\cr
\frac{\partial\,}{\partial\eta^k}D_\phi(p_\theta||p_\eta)\big|_{\eta=\theta}&=&0
\ee
and
\be
& &\hskip -2cm
\frac{\partial^2\,}{\partial\theta^k\partial\theta^l}D_\phi(p_\theta||p_\eta)\bigg|_{\eta=\theta}\cr
&=&\frac{\partial\,}{\partial\theta^k}
\int_{\Omega}{\rm d}\mu(x)\,\left[\ln_\phi\big(p_\theta(x)\big)-\ln_\phi\big(p_\eta(x)\big)\right]
\frac{\partial\,}{\partial\theta^l}p_\theta(x)\bigg|_{\eta=\theta}\cr
&=&
\int_{\Omega}{\rm d}\mu(x)\,\frac{1}{\phi\big(p_\theta(x)\big)}
\left[\frac{\partial\,}{\partial\theta^k}p_\theta(x)\right]
\left[\frac{\partial\,}{\partial\theta^l}p_\theta(x)\right]\cr
&=&\frac 1{Z(\theta)}g_{kl}(\theta).
\ee
Similar calculations give
\be
-\frac{\partial^2\,}{\partial\theta^k\partial\eta^l}D_\phi(p_\theta||p_\eta)\bigg|_{\eta=\theta}
=\frac{\partial^2\,}{\partial\eta^k\partial\eta^l}D_\phi(p_\theta||p_\eta)\bigg|_{\eta=\theta}
=\frac 1{Z(\theta)}g_{kl}(\theta).
\ee

\section{Information content}

In \cite {NJ02} the definition of deformed logarithm contains
the additional condition that the integral
\be
\int_1^0{\rm d}u\,\ln_\phi(u)=\int_0^1{\rm d}u\,\frac u{\phi(u)}<+\infty
\ee
converges. This condition is needed in the definition of entropy functional /
information content based on the deformed logarithm.
Introduce another strictly increasing positive function $\chi$ by
\be
\chi(v)=\left[
\int_0^{1/v}{\rm d}u\,\frac{u}{\phi(u)}
\right]^{-1}
\ee
The motivation for introducing this function comes from the fact that
it satisfies the following property.

\begin{lemma}
\label {lemma1}
\be
\frac{{\rm d}\,}{{\rm d}v}v\ln_\chi(1/v)
&=&-\ln_\phi(v) -\int_0^1{\rm d}u\,\frac{u}{\phi(u)}.
\ee
\end{lemma}

\begin{proof}
\be
\frac{{\rm d}\,}{{\rm d}v}v\ln_\chi(1/v)
&=&\ln_\chi(1/v)-\frac{1}{v\chi(1/v)}\cr
&=&\int_1^{1/v}{\rm d}u\,\frac{1}{\chi(u)}
-\frac{1}{v}\int_0^{v}{\rm d}u\,\frac{u}{\phi(u)}\cr
&=&\int_1^{1/v}{\rm d}u\,
\int_0^{1/u}{\rm d}z\,\frac{z}{\phi(z)}
-\frac{1}{v}\int_0^{v}{\rm d}u\,\frac{u}{\phi(u)}\cr
&=&-\int_1^v{\rm d}u\,\frac{1}{u^2}\int_0^{u}{\rm d}z\,\frac{z}{\phi(z)}
-\frac{1}{v}\int_0^{v}{\rm d}u\,\frac{u}{\phi(u)}\cr
&=&-\int_0^1{\rm d}z\,\frac{z}{\phi(z)}
-\ln_\phi(v),
\ee
which is the desired result.

\end{proof}

Define information content (also called entropy functional)
$I_\phi(p)$ of a pdf $p$ in ${\cal M}_1(\mu)$ by
\be
I_\phi(p)&=&\int_{\Omega}{\rm d}\mu(x)\,p(x)\ln_\chi(1/p(x))
\ee
whenever the integral converges. Using the lemma one verifies immediately
that $I_\phi(p)$ is a concave function of $p$.
A short calculation gives
\be
I_\phi(p)
&=&\int_{\Omega}{\rm d}\mu(x)\,p(x)\int_1^{1/p(x)}{\rm d}u\,\frac{1}{\chi(u)}\cr
&=&\int_{\Omega}{\rm d}\mu(x)\,p(x)\int_1^{p(x)}\frac{1}{\chi(1/v)}\,{\rm d}\frac{1}{v}\cr
&=&\int_{\Omega}{\rm d}\mu(x)\,p(x)\int_1^{p(x)}\left[\int_0^v{\rm d}u\,\frac{u}{\phi(u)}\right]
{\rm d}\frac{1}{v}\cr
&=&\int_{\Omega}{\rm d}\mu(x)\,p(x)\left[\frac{1}{p(x)\chi\big(1/p(x)\big)}
-\frac{1}{\chi(1)}-\ln\phi\big(p(x)\big)\right]\cr
&=&-\frac{1}{\chi(1)}-\int_{\Omega}{\rm d}\mu(x)\,\int_0^{p(x)}{\rm d}u\,\ln_\phi(u).
\ee
This implies that
\be
I_\phi(p)-I_\phi(p')=-\int_{\Omega}{\rm d}\mu(x)\,\int_{p'(x)}^{p(x)}{\rm d}u\,\ln_\phi(u),
\ee
and hence
\be
D_\phi(p||p')&=&I_\phi(p')-I_\phi(p)-\int_{\Omega}{\rm d}\mu(x)\,\big(p(x)-p'(x)\big)\ln_\phi\big(p'(x)\big).
\label {di}
\ee
This relation links the divergence $D_\phi(p||p')$ with the information function $I_\phi(p)$.

The following result shows that the $\phi$-exponential family is a conditional maximizer of $I_\phi$.
It also shows that the scale function $F$ is the Legendre transform of the information content $I_\phi$

\begin{theorem}
Let $\big(p_\theta\big)_{\theta\in D}$ be $\phi$-exponential, with estimator $c$
and scale functions $F$ and $G$. Then there exists a constant $F_0$ such that
\be
F(\theta)=F_0+\min_{p\in{\cal M}_1(\mu)}\{\Eo_p\theta^kc_k -I_\phi(p)\}.
\label {minfree}
\ee
The minimum is attained for $p=p_\theta$.
In particular, $F(\theta)$ is a convex function of $\theta$ and
$p_\theta$ maximizes $I_\phi(p)$ under the constraint that
\be
\Eo_p\theta^kc_k=\Eo_\theta\theta^kc_k.
\ee
\end{theorem}

\begin{proof}
Let us first show that for any pdf $p$
\be
\Eo_p\theta^kc_k -I_\phi(p)\ge \Eo_\theta\theta^kc_k- I_\phi(p_\theta).
\label {minfreeenerg}
\ee
One has
\be
& &\hskip -2cm
\int_{\Omega}{\rm d}\mu(x)\,\big(p(x)-p_\theta(x)\big)\ln_\phi\big(p_\theta(x)\big)\cr
&=&\int_{\Omega}{\rm d}\mu(x)\,\big(p(x)-p_\theta(x)\big)
\left[G(\theta)-\theta^kc_k\right]\cr
&=&-(\Eo_p-\Eo_\theta)\theta^kc_k.
\ee
Hence, (\ref {di}) becomes now
\be
D_\phi(p||p_\theta)=
I_\phi(p_\theta)-I_\phi(p)+(\Eo_p-\Eo_\theta)\theta^kc_k.
\ee
But one has always $D_\phi(p||p_\theta)\ge 0$. Therefore, (\ref {minfreeenerg}) follows.

Next calculate, using the lemma,
\be
\frac{\partial\,}{\partial\theta^k}I_\phi(p_\theta)
&=&\int{\rm d}\mu(x)\,
\left(-\ln_\phi\big(p_\theta(x)\big)-\int_0^1{\rm d}u\,\frac u{\phi(u)}\right)
\frac {\partial\,}{\partial\theta^k}p_\theta(x)\cr
&=&\int{\rm d}\mu(x)\,
\left(-G(\theta)+\theta^lc_l(x)-\int_0^1{\rm d}u\,\frac u{\phi(u)}\right)
\frac {\partial\,}{\partial\theta^k}p_\theta(x)\cr
&=&\int{\rm d}\mu(x)\,
\left(\theta^lc_l(x)\right)
\frac {\partial\,}{\partial\theta^k}p_\theta(x)\cr
&=&\frac {\partial\,}{\partial\theta^k}\left(\Eo_\theta\theta^lc_l\right)-\Eo_\theta c_k.
\ee
Because $c$ is an estimator with scale function $F$ one obtains
\be
\frac{\partial\,}{\partial\theta^k}\left(
\Eo_\theta\theta^lc_l
- I_\phi(p_\theta)\right)
&=&\frac{\partial\,}{\partial\theta^k}F(\theta).
\ee
Hence there exists a constant $F_0$ for which
\be
F(\theta)=F_0+\Eo_\theta\theta^lc_l
- I_\phi(p_\theta).
\label {dualrel}
\ee
In combination with (\ref {minfreeenerg}) this results in (\ref {minfree}).
\end{proof}

Without restriction one can assume $F_0=0$.
In statistical physics the function $F(\theta)$ is the free energy as a function of
temperature, up to some proportionality factor.

\paragraph {Example 4}
Let $\phi(u)=u^{2-q}/q$, with $0<q<2$. This is of course only a reparametrization of example 2,
which is done to recover expressions found in the literature. The deformed logarithm is
given by
\be
\ln_\phi(u)
&=&\frac q{q-1}(u^{q-1}-1)\qquad\hbox{ if }q\not=1\cr
&=&\log(u)\qquad\hbox{ if }q=1.
\ee
One obtains $\chi(v)=v^q$ and hence
\be
I_\phi(p)&=&\int{\rm d}\mu(x)\,p(x)\frac {1-p(x)^{q-1}}{q-1}.
\ee
This is the entropy functional proposed by Tsallis \cite {TC88}
as a basis for nonextensive thermostatistics, and reported earlier in
the literature by Havrda and Charvat \cite {HC67}
and by Dar\'oczy \cite {DZ70}. 
The corresponding expression for the divergence is
\be
D_\phi(p,p')
&=&\frac 1{q-1}
\int{\rm d}\mu(x)\,
p(x)\left[p(x)^{q-1}-p'(x)^{q-1}\right]\cr
& &
-\int{\rm d}\mu(x)\,\left[p(x)-p'(x)\right]p'(x)^{q-1}.
\ee

\section{Dual coordinates}

Introduce dual coordinates
\be
\eta_k=\Eo_\theta c_k=\frac{\partial F}{\partial\theta^k}.
\label {etadef}
\ee
Assume (\ref {escort}) holds. Then, one obtains from (\ref {Fderiv})
\be
\frac{\partial\eta_k}{\partial\theta^l}
&=&\frac{\partial\,}{\partial\theta^l}\Eo_\theta c_k\cr
&=&\frac{\partial^2\,}{\partial\theta^l\partial\theta^k}F(\theta)\cr
&=&-Z(\theta)
\left[\Fo_\theta c_kc_l-\big(\Fo_\theta c_k\big)\big(\Fo_\theta c_l\big)\right]\cr
&=&-\frac 1{Z(\theta)}g_{kl}(\theta).
\ee
To obtain the last line a $\phi$-exponential family has been assumed.
This relation implies
\be
\frac{\partial\theta^k}{\partial\eta_l}=-Z(\theta)g^{kl}(\theta).
\label {ttt}
\ee
Thes are the orthogonality relations between the two sets of coordinates $\theta$ and $\eta$.
Next we derive the dual relation of (\ref {etadef}).

\begin{proposition}
Let $\big(p_\theta\big)_{\theta\in D}$ be $\phi$-exponential.
Assume the regularity condition (\ref {reg}) is satisfied.
Then
\be
\theta^k=\frac{\partial\,}{\partial\eta_k}I_\phi(p_\theta).
\label {thetadef}
\ee
\end{proposition}

\begin{proof}
One calculates (assume integration and partial derivative can be interchanged),
using Lemma (\ref {lemma1}),
\be
\frac{\partial\,}{\partial \theta^k}I_\phi(p_\theta)
&=&-\int_{\Omega}{\rm d}\mu(x)\,\left[
\ln_\phi\big(p_\theta(x)\big)+\int_0^1{\rm d}u\,\frac{u}{\phi(u)}\right]
\frac{\partial\,}{\partial \theta^k}p_\theta(x)\cr
&=&-\int_{\Omega}{\rm d}\mu(x)\,\left[
G(\theta)-\theta^lc_l(x)
+\int_0^1{\rm d}u\,\frac{u}{\phi(u)}\right]
\frac{\partial\,}{\partial \theta^k}p_\theta(x)\cr
&=&\int_{\Omega}{\rm d}\mu(x)\,
\theta^lc_l(x)
\frac{\partial\,}{\partial \theta^k}p_\theta(x).
\ee
To obtain the last line the regularity condition has
been used.
Use now that $p_\theta$ satisfies (\ref {escort}). One obtains
\be
\frac{\partial\,}{\partial \theta^k}I_\phi(p_\theta)
&=&Z(\theta)\Fo_\theta\theta^lc_l(\Fo_\theta c_k-c_k)\cr
&=&-Z(\theta)\theta^lg_{lk}(\theta).
\ee
In combination with (\ref {ttt}) this gives
\be
\frac{\partial\,}{\partial\eta_l}I_\phi(p_\theta)
&=&\left(\frac{\partial\,}{\partial\theta^l}I_\phi(p_\theta)\right)
\frac{\partial\theta^k}{\partial\eta_l}\cr
&=&\left(-Z(\theta)\theta^mg_{ml}(\theta)\right)
\left(-\frac{1}{Z(\theta)}g^{kl}(\theta)
\right)\cr
&=&\theta^l.
\ee

\end{proof}

Equation (\ref {thetadef}) is the dual relation of (\ref {etadef}). 
Expression (\ref {dualrel}) can now be written as
\be
F(\theta)+E(\eta)=\theta^k\eta_k
\ee
with $E(\eta)=I_\phi(p_\theta)$.

\section{Discussion}

The present paper introduces generalized exponential families, and
calls them $\phi$-exponential because they depend on the choice of a strictly
positive non-decreasing function $\phi$ of $(0,+\infty)$.
Several properties, known to hold for the exponential family,
can be generalized. The paper starts with a generalization of the well-known lower
bound of Cramer and Rao, involving the concept of escort probability
distributions. It is shown that the $\phi$-exponential family optimizes
this generalized lower bound. The metric tensor, which generalizes the
Fisher information, depends on both the family of pdfs and the escort
family, and determines the geometry of the statistical manifold.

In the final part of the paper deals with the dual structure of the statistical manifold.
In statistical physics this duality
maps mean energy onto temperature and entropy onto free energy.
The dual structure is shown to exist in the general
context of $\phi$-exponential families.

Throughout the paper the number of parameters $n$ has been assumed to be finite.
A non-parametrized approach to statistical manifolds is found in \cite {PS95}.
The extension of the present work to this more abstract context has
not been considered.

\section*{Acknowledgments}
I am thankful to S. Abe who urged me to study the geometry
of statistical distributions. I thank Dr. Ch. Vignat for pointing
out ref. \cite {PS95}.


\end{document}